\documentclass{pasj02}
\usepackage[switch,mathlines]{lineno}
\Received{$\langle$reception date$\rangle$}
\Accepted{$\langle$acception date$\rangle$}
\Published{$\langle$publication date$\rangle$}
\usepackage{natbib}
\begin{document}

\title{Three-dimensional core-collapse supernova models with phenomenological treatment of neutrino flavor conversions}
\author{Kanji \textsc{Mori},\altaffilmark{1,}\footnotemark[*] Tomoya \textsc{Takiwaki},\altaffilmark{1} Kei \textsc{Kotake},\altaffilmark{2,3} and Shunsaku \textsc{Horiuchi}\altaffilmark{4,5}}%
\altaffiltext{1}{National Astronomical Observatory of Japan, 
 2-21-1 Osawa, Mitaka, Tokyo 181-8588, Japan}
 \altaffiltext{2}{Department of Applied Physics, Faculty of Science, Fukuoka University, 8-19-1 Nanakuma, Jonan-ku, Fukuoka-shi, Fukuoka 814-0180, Japan}
\altaffiltext{3}{Institute for Theoretical Physics, University of Wroc\l aw, 50-204 Wroc\l aw, Poland}
\altaffiltext{4}{Center for Neutrino Physics, Department of Physics, Virginia Tech, Blacksburg, VA 24061, USA}
\altaffiltext{5}{Kavli IPMU (WPI), UTIAS, The University of Tokyo, Kashiwa, Chiba 277-8583, Japan}

\email{kanji.mori@nao.ac.jp}


\maketitle

\begin{abstract}

We perform three-dimensional supernova simulations with a phenomenological treatment of neutrino flavor conversions. We show that the explosion energy can increase to as high as $\sim10^{51}$\,erg depending on the critical density for the onset of flavor conversions, due to a significant enhancement of the mean energy of electron antineutrinos. Our results confirm previous studies showing such energetic explosions, but for the first time in three-dimensional configurations. 
In addition, we predict neutrino and gravitational wave (GW) signals from a nearby supernova explosion aided by  flavor conversions. We find that the neutrino event number decreases because of the reduced flux of heavy-lepton neutrinos. In order to detect GWs,
next-generation GW telescopes such as Cosmic Explorer and Einstein Telescope are needed even if the supernova event is located at the Galactic center. These findings show that the neutrino flavor conversions can significantly change supernova dynamics and highlight the importance of further studies on the quantum kinetic equations to determine the conditions of the conversions and their asymptotic states.

\end{abstract}

 \section{Introduction}

  \begin{figure*}
\begin{center}
 \includegraphics[width=80mm]{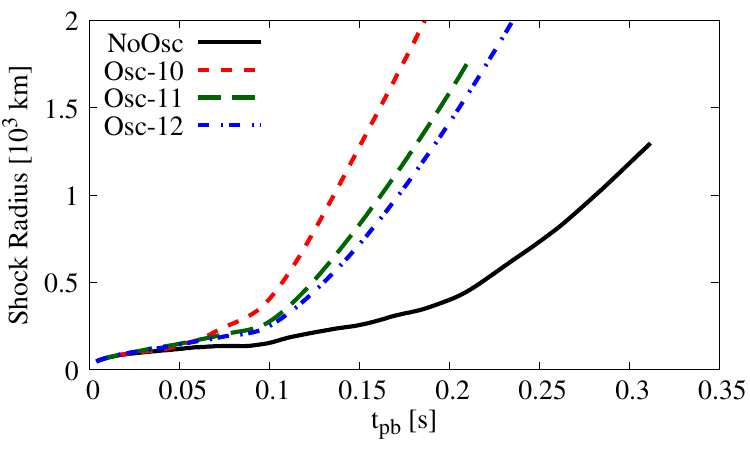}
  \includegraphics[width=80mm]{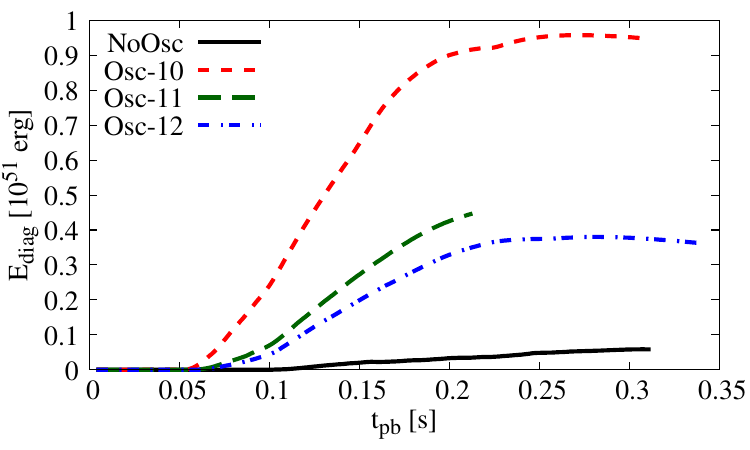}
\end{center}
\caption{The angular-averaged radius of the bounce shock (left) and the diagnostic explosion energy (right) as a function of the post-bounce time $t_\mathrm{pb}$. }\label{fig:rsh}
\end{figure*}

Since the paradigm of the neutrino-driven delayed explosion mechanism for core-collapse supernovae was proposed \citep{1985ApJ...295...14B}, substantial efforts have been dedicated to develop multi-dimensional supernova models coupled with the neutrino transport \citep[e.g.][]{2011ApJ...738..165S,2014ApJ...786...83T,2015ApJ...807L..31L,2015ApJ...808L..42M,2015ApJ...801L..24M,2016ApJ...817...72P,2019JPhG...46a4001P,2017MNRAS.472..491M,2018ApJ...865...81O,2018ApJ...855L...3O,2020MNRAS.491.2715B,2024ApJ...964L..16B,2020ApJ...896..102K,2021ApJ...915...28B,2024arXiv240508367N}. Various numerical methods have been developed to solve the classical neutrino transport accurately enough to make reliable prediction. This is critical because the neutrino heating rate behind the bounce shock is essential for the explosion \citep[e.g.][]{2002A&A...396..361R,2009ApJ...698.1174L,2012ApJS..199...17S}. However, the effect of neutrino flavor oscillations has not been fully considered in these studies. 

The neutrino oscillations in dense neutrino gases have been investigated by solving the quantum kinetic equation in simplified situations that mimic supernova backgrounds. Such studies revealed that there exist several flavor instabilities that can happen when certain conditions are satisfied. An example is the fast flavor instability (FFI) \citep{2005PhRvD..72d5003S,2009PhRvD..79j5003S}, which grows if and only if the electron lepton number crosses zero \citep{2022PhRvL.128h1102D,2022PhRvD.105j1301M}. More recently, the collisional flavor instability (CFI) induced by the difference in the interaction rates for $\nu_e$ and $\bar{\nu}_e$ was discovered \citep{2023PhRvL.130s1001J}. Many authors have adopted the post-process technique to determine the asymptotic state for the instabilities and whether they appear in supernovae \citep{2019PhRvD.100d3004A,2021PhRvD.103f3033A,2020PhRvD.101d3009J,2020PhLB..80035088M,2021PhRvD.104b3011S,2021PhRvD.104h3025N,2021PhRvD.103h3013R,2021PhRvD.104j3003W,2022PhRvD.106l3013K,2022PTEP.2022g3E01S,2022ApJ...924..109H,2023PhRvD.107l3021Z,2023PhRvD.107h3034L,2023PhRvD.108l3024L,2023PhRvD.108h3002X,2023PhRvD.107h3016X,2023PhRvD.107d3024F,2024PhRvD.109b3012A,2024arXiv240204741D}.

It is extremely challenging to couple neutrino oscillations with supernova hydrodynamics, because the time scale for the oscillations is much shorter than the time steps in hydrodynamical simulations. However, several effective methods have been proposed. For example, \citet{2023PhRvL.130u1401N} introduced an artificial attenuation parameter in the quantum kinetic equation to reduce the required computational resources. In addition, \citet{2023PhRvL.131f1401E,2023PhRvD.107j3034E} developed a phenomenological approach to implement neutrino oscillations, assuming flavor equipartition. By performing one- and two-dimensional simulations, the authors discovered the oscillations to modify many quantities, e.g., shock revival and explosion energies. However, the outcomes of core collapse are also strongly dependent on the dimensionality of simulations, and a three-dimensional treatment is essential to make reliable prediction of signals \citep[e.g.][]{2019MNRAS.487.1178P,2021MNRAS.508..966T,2023PhRvD.107j3015V}. Therefore, three-dimensional simulations incorporating phenomenological inclusion of neutrino oscillations are warranted. 

In this study, we perform three-dimensional simulations to reveal the impact of the neutrino flavor conversions. As shown in the right panel of Figure~\ref{fig:rsh}, we find that the flavor conversions can render explosion energies as high as $\sim10^{51}$ erg. In addition, we predict impacts of the flavor conversions on gravitational wave (GW) and neutrino signals from a nearby supernova event.

 \section{Method}

We perform three-dimensional stellar core-collapse simulations with the \texttt{3DnSNe} code \citep{2016MNRAS.461L.112T}, adopting a non-rotating $11.2M_\odot$ progenitor model with the solar metallicity in \cite{2002RvMP...74.1015W}.  
The code adopts the three-flavor isotropic diffusion source approximation \citep[IDSA;][]{2009ApJ...698.1174L,2014ApJ...786...83T,2018ApJ...853..170K} to solve the neutrino transport. In order to implement the neutrino flavor conversions, we follow the phenomenological method developed in \citet{2023PhRvL.131f1401E,2023PhRvD.107j3034E}. They simply  assume that flavor equipartition is achieved within a shorter timescale than the simulation time step, because numerical treatments of asymptotic state for the conversions 
in global simulations are still under debate \cite[see][for discussions]{Dasgupta2018,Bhattacharyya2020,Bhattacharyya2021,Zaizen2023a,2023PhRvD.107l3021Z,Xiong2023b}.
We introduce a critical density $\rho_\mathrm{c}$ as a free parameter, below which equations (9a)--(9c) in \citet{2023PhRvD.107j3034E} are applied to the neutrino distribution functions. From these equations, the lepton number conservation is guaranteed as in \cite{2023PhRvD.107j3034E}. In our simulations, we assume $\rho_\mathrm{c}=10^{10}$, $10^{11}$, and $10^{12}$\,g cm$^{-3}$. This choice can be justified by previous works \citep{2023PhRvD.108l3024L,2024PhRvD.109b3012A} which report that by analyzing the quantum kinetic equation in supernova backgrounds, CFI appears around this density. We also develop a model without neutrino oscillations for comparison. We summarize the models  in this study in Table 1. 

\begin{table}
\tbl{The summary of our models. The first column shows the model name, $\rho_\mathrm{c}$ is the critical density for the onset of neutrino flavor equipartition, $t_{\mathrm{pb},\;1000}$ is the post-bounce time at which the bounce shock reaches the radius of 1000\,km, $E_\mathrm{diag}$ is the diagnostic explosion energy, $M_\mathrm{Ni}$ is the ejected nickel mass, and $M_\mathrm{PNS}$ is the proto-neutron star mass. The last three quantities are evaluated at $t_\mathrm{pb}=t_{\mathrm{pb},\;1000}$.}{
\begin{tabular}{cccccc}
\hline
\multicolumn{1}{c}{} & $\rho_\mathrm{c}$ &  $t_\mathrm{pb,\;1000}$& $E_\mathrm{diag}$ &$M_\mathrm{Ni}$&$M_\mathrm{PNS} $ \\
 & [g cm$^{-3}]$ & [ms] & $[10^{51}$\,erg] & [$M_\odot$] & [$M_\odot$] \\
\hline
\texttt{NoOsc}  & 0 & 282 & 0.063 & 0.0098 & 1.33 \\
\texttt{Osc-10}  & $10^{10}$ & 136 & 0.54 & 0.061 & 1.25  \\
\texttt{Osc-11}  & $10^{11}$ & 163 & 0.31 & 0.045 & 1.28 \\
\texttt{Osc-12}  & $10^{12}$ & 172 & 0.26 & 0.038 & 1.29 \\
\hline
\end{tabular}} \label{tab:first}
\end{table}

Most of our simulations were run until core post-bounce time $t_\mathrm{pb}\sim0.3$\,s, but the model with $\rho_\mathrm{c}=10^{11}$\,g cm$^{-3}$ (\texttt{Osc-11} Model) was terminated at $t_\mathrm{pb}\sim0.2$\,s, due to it showing early signs of behaviors intermediate between the \texttt{Osc-10} and \texttt{Osc-12} Models.

 \begin{figure*}
\begin{center}
 \includegraphics[width=140mm]{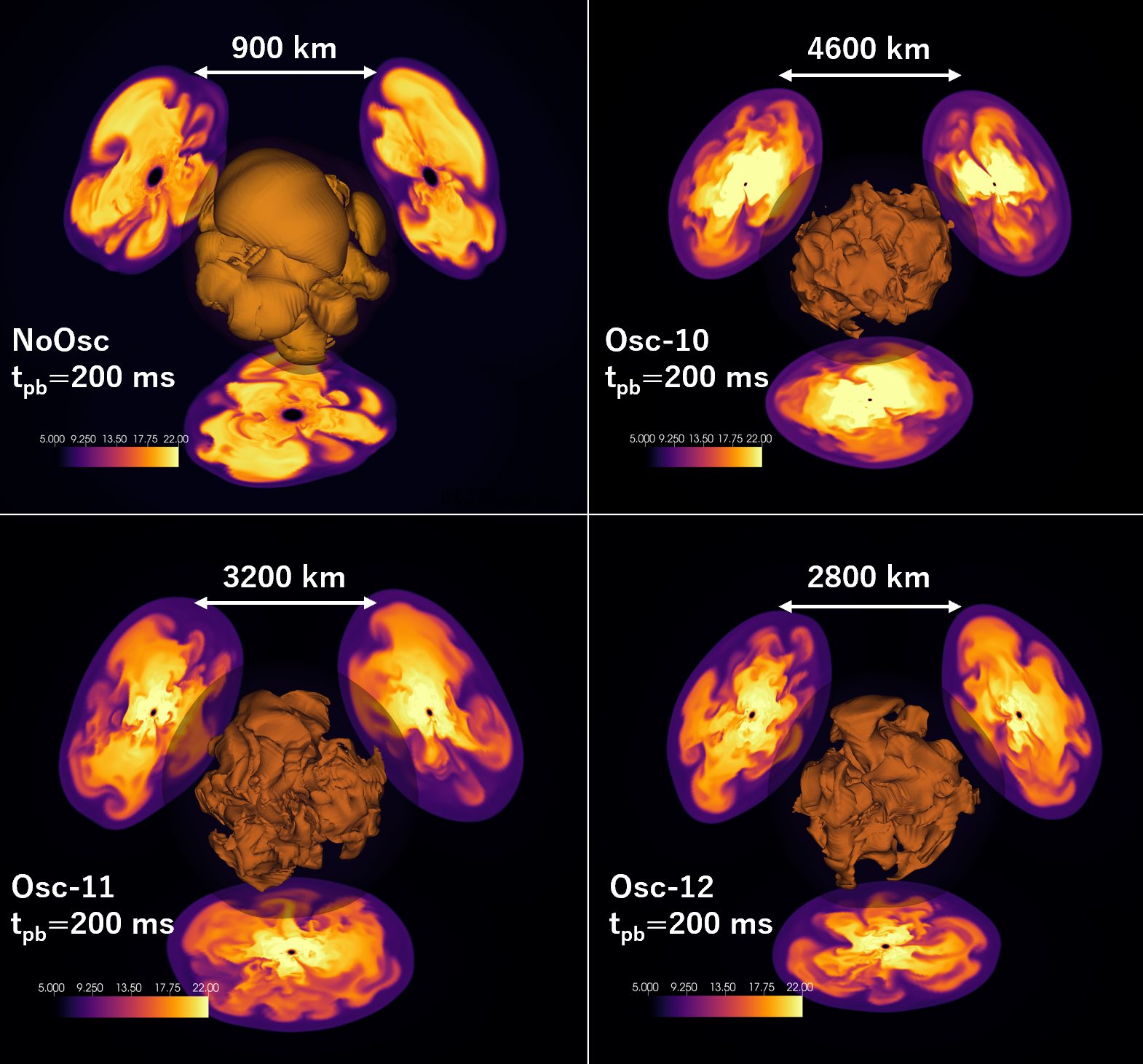}
\end{center}
\caption{Snapshots at $t_\mathrm{pb}=0.2$\,s for our 4 models. The panels indicate entropy isosurfaces of $s/k_\mathrm{B}=7$ and $17$ per baryon. Note that the spatial scale is different for each panel.}\label{fig:snapshot}
\end{figure*}

 \section{Results}

Figure~\ref{fig:rsh} shows the radius of the bounce shock as a function of the post-bounce time $t_\mathrm{pb}$. In all of our models, the bounce shock first stalls at $t_{\mathrm{pb}}\lesssim0.05$\,s before reviving in later phases. We can see that the propagation of the shock wave is fastest in the \texttt{Osc-10} Model and slowest in the \texttt{NoOsc} Model. This trend is similar to two-dimensional models where the same phenomenological FFI treatment was developed \citep{2023PhRvL.131f1401E}.

 The effect of the flavor conversion can be found in the time snapshots at $t_\mathrm
{pb}=0.2$\,s for our models shown in Figure~\ref{fig:snapshot} as well. The figure shows that the neutrino-heated material extends to $r\sim1000$--3000\,km when the flavor conversion is considered, while it stays within $r\sim500$\,km for the \texttt{NoOsc} Model. One can also see that the peak entropy reaches $s/k_\mathrm{B}=25$ per baryon for the \texttt{Osc-10} Model, whereas it is as low as 18 per baryon for the \texttt{NoOsc} Model, where  $k_\mathrm{B}$ is the Boltzmann constant. This result implies that the flavor conversion enhances the neutrino heating and thus helps the explosion.

Figure~\ref{fig:rsh} also shows the diagnostic explosion energy $E_\mathrm{diag}$ of each model, defined as
\begin{eqnarray}
    E_\mathrm{diag}=\int_D dV\left(\frac{1}{2}\rho v^2+e-\rho\Phi\right),\label{Ediag}
\end{eqnarray}
where  $\rho$ is the density, $v$ is the velocity, $e$ is the internal energy, $\Phi$ is the gravitational potential, and $D$ is the region where the integrand is positive and the velocity is outward. In our \texttt{NoOsc} Model, $E_\mathrm{diag}$ does not reach $0.1\times10^{51}$\,erg at the end of our simulation at $t_\mathrm{pb}\sim0.3$\,s, similar to reported in \citet{2015MNRAS.453..287M} and \citet{2016MNRAS.461L.112T} for the same progenitor. On the other hand, the explosion energy is significantly enhanced when neutrino flavor conversions are included. The explosion energy is highest in the \texttt{Osc-10} Model, where $0.96\times10^{51}$\,erg is reached; this is followed by the \texttt{Osc-11} Model and finally the \texttt{Osc-12} which is still more energetic than the \texttt{NoOsc} Model. 

The enhancement in the explosion energy has important observational implications. Supernova surveys have shown that the typical explosion energy for observed supernova events is $\sim0.6\times10^{51}$\,erg \citep{2022A&A...660A..41M}, which is higher than $E_\mathrm{diag}$ in \texttt{NoOsc} Model. Our results imply that the neutrino oscillations could play an essential role to reproduce the observed supernova events. We note, however, that recent long-term three-dimensional simulations in \citet{2021ApJ...915...28B} and \citet{2024ApJ...964L..16B} indicate that the explosion energy could continue to increase at $t_\mathrm{pb}>1$\,s. It is therefore desirable to perform long-term simulations coupled with the flavor conversions to determine asymptotic values of the explosion energy. Nevertheless, our \texttt{Osc-10} Model already exceeds the typical observed explosion energy even without such late-time growth. 

 \begin{figure*}
\begin{center}
 \includegraphics[width=160mm]{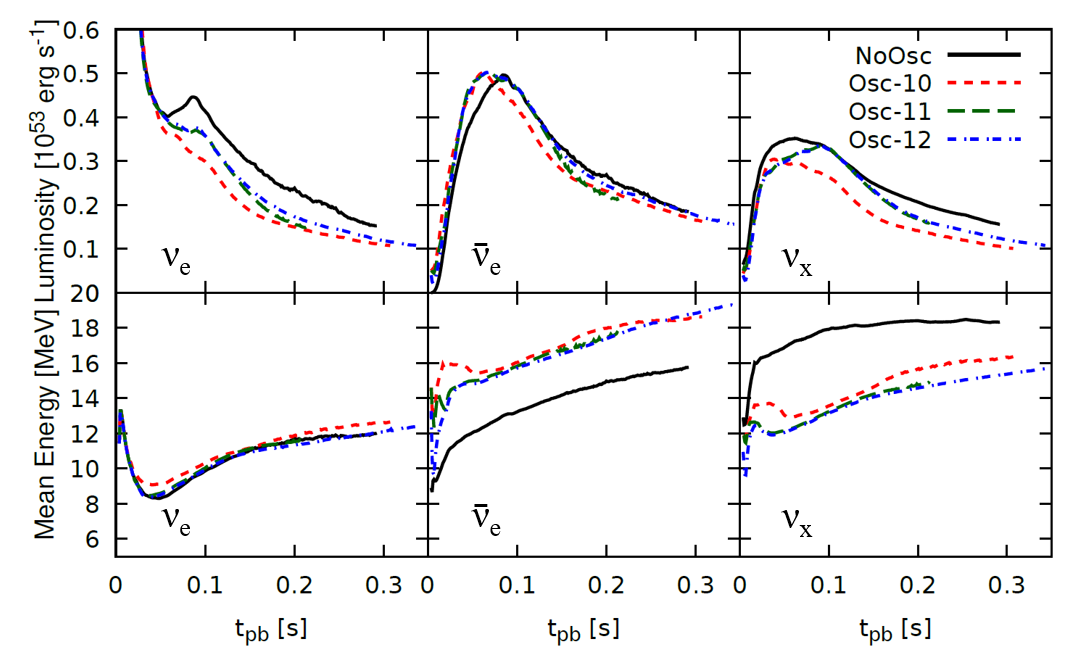}
\end{center}
\caption{The neutrino luminosity (upper) and mean energy (lower) measured at $r=500$\,km as functions of time, for electron, anti-electron, and heavy lepton neutrinos from left to right as labeled. }\label{fig:neu}
\end{figure*}

 \begin{figure}
\begin{center}
 \includegraphics[width=80mm]{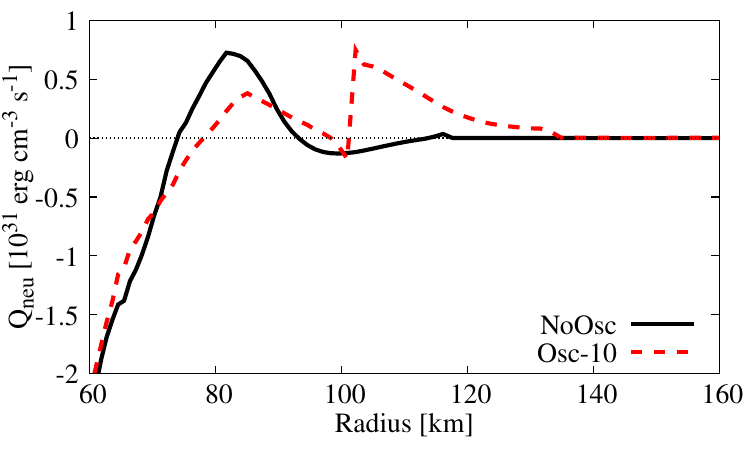}
\end{center}
\caption{The radial profile for the net neutrino heating rate $Q_\mathrm{neu}$ at $t_\mathrm{pb}=0.05$\,s. The critical density $\rho_\mathrm{c}=10^{10}$\,g cm$^{-3}$ for \texttt{Osc-10} Model is located at $r\approx100$\,km.}\label{fig:Qneu}
\end{figure}

The impact on the explosion energy can be attributed to differences in the neutrino mean energies. Figure~\ref{fig:neu} shows the luminosity and mean energy for each neutrino flavor. One can see that the flavor equipartition suppresses the $\nu_x$ mean energy and enhances the $\bar{\nu}_e$ mean energy. This change has a significant impact on the neutrino heating. Figure~\ref{fig:Qneu} shows the net neutrino heating rate $Q_\mathrm{neu}$ at $t_\mathrm{pb}=0.05$\,s for the \texttt{NoOsc} and \texttt{Osc-10} Models. As one can see, the heating rate in the gain layer is increased for the \texttt{Osc-10} Model. This is explained by the fact that the heating rate is proportional to $L_\nu\langle\epsilon_\nu\rangle^2$, where $L_\nu$ and $\langle\epsilon_\nu\rangle$ are the electron (anti)neutrino luminosity and mean energy, respectively.

As shown in Fig.~\ref{fig:neu}, the neutrino luminosities are affected by the flavor conversions. At $t_\mathrm{pb}\lesssim0.05$\,s, the $\bar{\nu}_e$ ($\nu_x$) luminosity for the models with the flavor conversions becomes higher (lower) than the one for the \texttt{NoOsc} Model, because of the $\nu_x\rightarrow\nu_e$, $\bar{\nu}_e$ conversion. In the later phase, the luminosities for all the flavors become smaller than the ones for the \texttt{NoOsc} Model because of the suppressed mass accretion. On the other hand, the decrease in the $\nu_x$ mean energy and the increase in the $\bar{\nu}_e$ mean energy are attributed to the $\nu_x\rightarrow\nu_e$, $\bar{\nu}_e$ conversion. Since $\nu_x$ has the highest mean energy on the neutrinosphere, the flavor equipartition can redistribute the $\nu_x$ energy to the other flavors to enhance their mean energies. We note that the effect on the neutrino mean energies was also reported in the \texttt{M11.2\_2D} Models in \cite{2023PhRvL.131f1401E}. 

Table 1 shows $E_\mathrm{diag}$, the ejected nickel mass $M_\mathrm{Ni}$, and the proto-neutron star mass $M_\mathrm{PNS}$ for each model. One can see that the neutrino flavor conversions significantly increase $M_\mathrm{Ni}$ because of the enhanced neutrino heating rate. At the same time, $M_\mathrm{PNS}$ decreases because mass accretion is suppressed.     

\begin{table}
\tbl{The expected neutrino event number $N_\nu$ and the positron mean energy $\langle E_{e^+}\rangle$ observed by HK during the $t_\mathrm{pb}=0$--$0.20$\,s time window from a nearby supernova event at $D=8.5$\,kpc. The ratios of $N_\nu$, normalized to the \texttt{NoOSc} Model, are also shown to highlight the differences.}{
\begin{tabular}{ccccccc}
\hline
&\multicolumn{3}{c}{NH}&\multicolumn{3}{c}{IH}\\
 & $N_\nu$ & Ratio & $\langle E_{e^+}\rangle$ & $N_\nu$ & Ratio &$\langle E_{e^+}\rangle$  \\
&&&[MeV]&&&[MeV]\\
\hline
 \texttt{NoOsc}  & 10117& 1 &24.8 & 12946 &1  &34.5  \\
 \texttt{Osc-10}  & 10528 & 1.04 & 25.2 & 6679 & 0.52 &21.9\\
 \texttt{Osc-11}  & 9859 & 0.97 & 21.0 & 6817 & 0.53 &18.7\\
\texttt{Osc-12}  & 9840 & 0.97 &20.7 &6716& 0.52 &18.5 \\
\hline
\end{tabular}} \label{tab:second}
\end{table}

The changes in $L_\nu$ and $\langle\epsilon_\nu\rangle$ due to oscillations affects neutrino observables from a nearby supernova event. Table 2 shows the neutrino event number observed by Hyper-Kamiokande \citep[HK;][]{2021ApJ...916...15A} during the first $t_\mathrm{pb}=0$--$0.20$\,s. The event rate is estimated as
 \begin{eqnarray}
     \frac{dN_\nu}{dt}=N_\mathrm{tar}\int_{E_\mathrm{th}}^\infty F(E)\sigma(E) dE,
\end{eqnarray}
where $N_\mathrm{tar}$ is the number of target protons in HK \citep{2018arXiv180504163H}, $F(E)$ is the $\bar{\nu}_e$ number flux, $\sigma(E)$ is the inverse $\beta$-decay cross section, and $E_\mathrm{th}=8.3$\,MeV \citep{2017ApJ...848...48K} is the threshold energy. We include the Mikheyev-Smirnov-Wolfenstein (MSW) effect \citep{1979PhRvD..20.2634W,1986NCimC...9...17M,PhysRevLett.56.1305} to obtain the $\bar{\nu}_e$ flux observed on Earth, $F_{\bar{\nu}_e}=pF^0_{\bar{\nu}_e}+(1-p)F^0_{\bar{\nu}_X}$, where $F^0$ is the flux before passing the MSW resonance, $p\approx0.676$ for the normal mass hierarchy (NH) and $p\approx0.0234$ for the inverted hierarchy (IH) \citep{2014ChPhC..38i0001O}. Although we assume that the flavor equipartition is achieved before neutrinos arrive at the MSW radius, neutrinos are still affected by the resonance. This is because the $\bar{\nu}_X$ spectrum is equal to that for either $\nu_e$ or $\bar{\nu}_e$, depending on the neutrino energy. 

We can find that, in the IH case, the neutrino event number $N_\nu$ decreases if the flavor conversions are considered, whereas it is almost unchanged in the NH case. The primary reason for the reduction in $N_\nu$ in the IH case is the reduction in the $\nu_X$ the mean energy.  It is also notable that, in the IH case, the positron mean energy $\langle E_{e^+}\rangle$ observed by HK is significantly decreased by the flavor conversions. This is also attributed to the reduced $\nu_X$ mean energy shown in Fig.~\ref{fig:neu}. Interestingly, it has recently been reported that modern theoretical models overestimate the mean energy of Kamiokande neutrino events from SN 1987A \citep{2023PhRvD.108h3040F,2024PhRvD.109h3025L}. In particular, this appears despite SN1987A showing average explosion energies. Flavor conversions reduce $\langle E_{e^+}\rangle$ without reductions in the explosion energy, which can mitigate the discrepancy between model prediction and SN 1987A observations. We note that \citet{2023PhRvD.108l3003N}  reported that the neutrino event rate is increased by the FFI in the IR case on the basis of the quantum kinetic equation. As discussed in the literature, this difference could be attributed to our assumption of $\nu_x=\bar{\nu}_x$.

 \begin{figure}
\begin{center}
 \includegraphics[width=80mm]{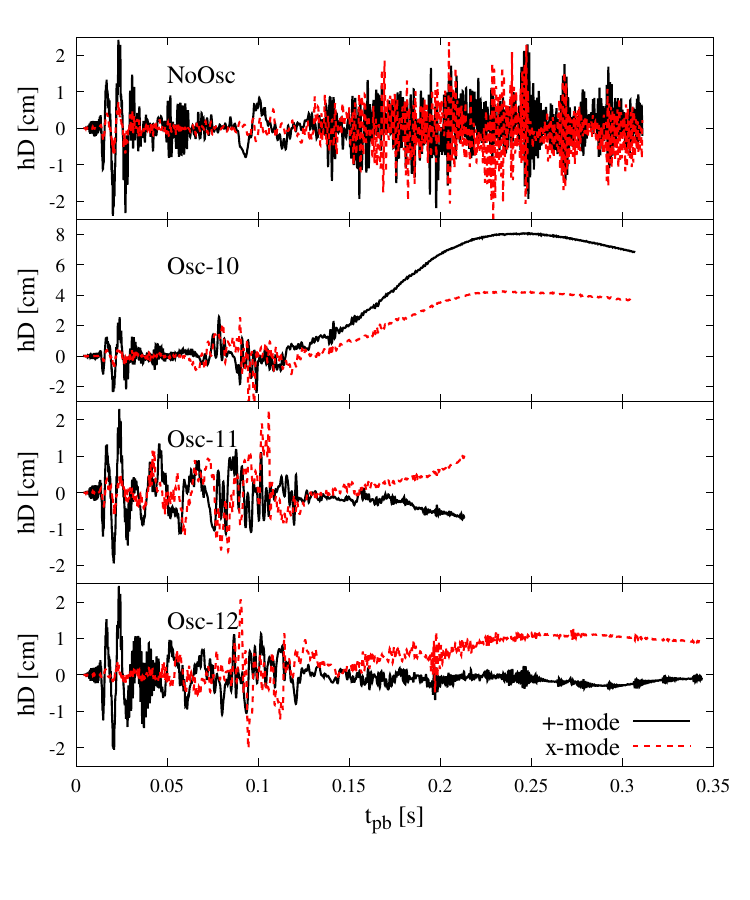}
\end{center}
\caption{The GW strain $h$ times the distance $D=8.5$\,kpc to the supernova event as a function of time. The observer is assumed to be located in the equatorial direction.  The solid curves show the plus-mode and the broken curves show the cross-mode. Note that the scale for the \texttt{Osc-10} Model is different from the others.}\label{fig:GW}
\end{figure}

Figure~\ref{fig:GW} shows the matter-generated GW strain $h$ from a supernova event at $D=8.5$\,kpc away. It is seen that the GW amplitude during the early phase $t_\mathrm{pb}\lesssim0.1$\,s, which is powered by the prompt convection in the protoneutron star, is similar among all models. However, at later phases,  the strain deviates from $h\sim0$  if the neutrino flavor conversion is considered. This ``memory'' effect can be attributed to the prolate morphology \citep[see, e.g.,][]{2009ApJ...707.1173M} and a high explosion energy for the \texttt{Osc-10} Model.

\begin{figure}
\begin{center}
 \includegraphics[width=80mm]{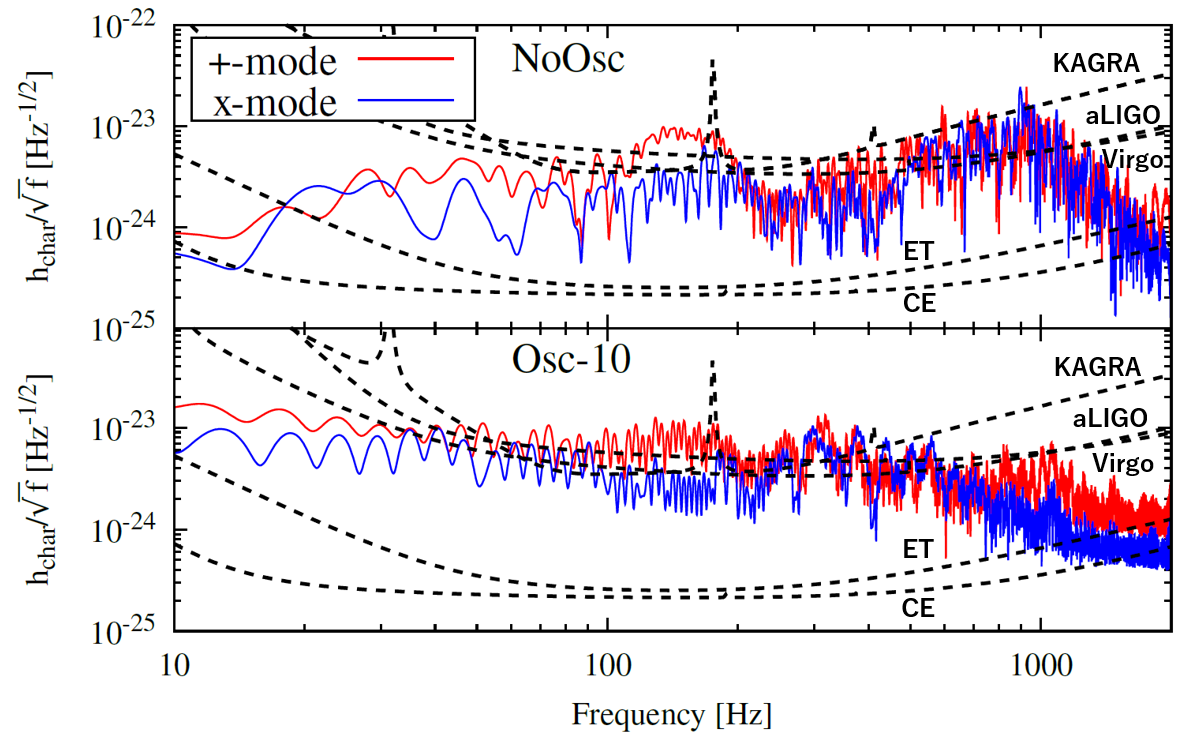}
\end{center}
\caption{The GW spectrum from a supernova event at $D=8.5$\,kpc away for the \texttt{NoOsc} and \texttt{Osc-10} Models. The sensitivity curves for Advanced LIGO, Advanced Virgo, KAGRA, ET, and CE are also shown.}\label{fig:spec}
\end{figure}

Figure~\ref{fig:spec} shows the GW spectrum 
\begin{eqnarray}
    h_\mathrm{char}=\sqrt{\frac{2G}{\pi c^3D^2}\frac{dE_\mathrm{GW}}{df}},
\end{eqnarray}
where $dE_\mathrm{GW}/df$ is the GW spectral energy density, for the \texttt{NoOsc} and \texttt{Osc-10} Models. One can find that a peak at $f\sim1$\,kHz for \texttt{NoOsc} Model vanishes for the \texttt{Osc-10} Model because the mass accretion on the protoneutron star is suppressed. It is also found that a low-frequency component at $f\sim10$\,Hz is higher for \texttt{Osc-10} Model because of the ``memory'' effect \citep[][]{2009ApJ...707.1173M} induced by the prolate morphology.  For both models, the GW signals are below the sensitivity curves for current GW detectors LIGO \citep{2015CQGra..32g4001L}, Virgo \citep{2015CQGra..32b4001A}, and KAGRA \citep{2021PTEP.2021eA101A}, if the supernova event is located at $D=8.5$\,kpc. However, the signals can be a target of next-generation detectors including Einstein Telescope \citep[ET;][]{ET} and Cosmic Explorer \citep[CE;][]{CE}, whose sensitivity is expected to reach $h_\mathrm{char}\sim10^{-24}$--$10^{-25}$.

\section{Conclusions}

In this study, we performed three-dimensional supernova simulations of the $11.2M_\odot$ progenitor taking in account neutrino flavor conversions in a phenomenological way following the method developed in \citet{2023PhRvL.131f1401E,2023PhRvD.107j3034E}. Specifically, we assumed that flavor equipartition is realized when the density is lower than a critical value $\rho_\mathrm{c}$. It is found that the $\nu_x\rightarrow\nu_e$,\,$\bar{\nu}_e$ enhances the mean energy of $\bar{\nu}_e$ and, as a result, the explosion energy is increased. In particular, when $\rho_c=10^{10}$\,g cm$^{-3}$, the explosion energy reaches $\sim10^{51}$\,erg at $t_\mathrm{pb}\sim0.3$\,s, which is an order of magnitude larger than the no-oscillation case and can even exceed the observed value for typical supernova events. In addition, the $\bar{\nu}_e$ event number observed by water-Cherenkov detectors decreases, because the $\nu_x$ luminosity and the mean energy are suppressed. We also find that the low-frequency GW signal is developed when $\rho_\mathrm{c}=10^{10}$\,g cm$^{-3}$. However, it would be difficult to detect GW signals from a supernova event at the Galactic center with the current GW detectors, regardless of $\rho_\mathrm{c}$ \citep{2016PhRvD..94j2001A}.

Our results indicate that the neutrino flavor conversion could significantly affect supernova dynamics and multi-messenger signals from nearby events. However, the condition for the flavor conversions and their asymptotic state are still under debate. It is desirable to investigate these issues on the basis of the quantum kinetic equation to go beyond the phenomenological treatment adopted here. Recent studies on the CFI showed that oscillations can occur in the region of $\rho=10^{10}$--$10^{12}$\,g cm$^{-3}$, regardless of progenitors and spatial dimensionality \citep{2023PhRvD.108l3024L,2024PhRvD.109b3012A}. From this viewpoint, our results could be interpreted as models with CFI, although its asymptotic state may not be full flavor equipartition \citep{2024PhRvD.109j3009K}. 

We note that IDSA, the neutrino transport treatment adopted in our simulations, adopts a ray-by-ray approximation. However, the information on the angular distribution is necessary to determine the unstable region for FFI. Therefore, it would be worthwhile to adopt transport schemes that retain the angular distribution, including the $S_N$ scheme \citep{2012ApJS..199...17S}.

Since three-dimensional simulations are computationally expensive, we focused only on the $11.2M_\odot$ progenitor model. However, \citet{2023PhRvL.131f1401E} found that the effects of the flavor equipartition depend on the progenitor mass. When they adopted a $20M_\odot$ progenitor, the explosion was rather hindered in their two-dimensional models. Aside from the progenitor dependence, the flavor conversions would affect other observables such as explosive nucleosynthesis \citep{2023MNRAS.519.2623F}, the neutron star kick \citep{2024PhRvD.109j3017N}, and the mass functions of black holes and neutron stars. Going forwards, it is therefore desirable to perform three-dimensional simulations systematically. 

\begin{ack}
The authors thank Alessandro Mirizzi and Jacob Ehring for stimulating discussions. Numerical computations were  carried out on Cray XC50 at the Center for Computational Astrophysics, National Astronomical Observatory of Japan. This work is supported by JSPS KAKENHI Grant Numbers  JP23K20848, JP23KJ2147, JP23K03400, JP23K22494, JP23K13107, JP23K25895, and JP24K00631, World Premier International Research Center Initiative (WPI Initiative), MEXT, Japan, and funding from Fukuoka University (Grant No.GR2302). This research is also supported by MEXT as “Program for Promoting researches on the Supercomputer Fugaku” (Structure and Evolution of the Universe Unraveled by Fusion of Simulation and AI; Grant Number JPMXP1020230406) and JICFuS. The work of SH is supported by the U.S.~Department of Energy Office of Science under award number DE-SC0020262, NSF Grant No.~PHY-2209420, and JSPS KAKENHI Grant Number JP22K03630 and JP23H04899. 
\end{ack}

\bibliography{ref} 
\bibliographystyle{aasjournal} 

\end{document}